# UCNN: A Convolutional Strategy on Unstructured Mesh

Mengfei Xu[1], Shufang Song[1], Xuxiang Sun[1], Weiwei Zhang[1]

**Abstract**    In machine learning for fluid mechanics, fully-connected neural network (FNN) only uses the local features for modelling, while the convolutional neural network (CNN) cannot be applied to data on structured/unstructured mesh. In order to overcome the limitations of FNN and CNN, the unstructured convolutional neural network (UCNN) is proposed, which aggregates and effectively exploits the features of neighbour nodes through the weight function. Adjoint vector modelling is taken as the task to study the performance of UCNN. The mapping function from flow-field features to adjoint vector is constructed through efficient parallel implementation on GPU. The modelling capability of UCNN is compared with that of FNN on validation set and in aerodynamic shape optimization at test case. The influence of mesh changing on the modelling capability of UCNN is further studied. The results indicate that UCNN is more accurate in modelling process.

✉ Shufang Song

shufangsong@nwpu.edu.cn

[1] Northwestern Polytechnical University, Xi 'an, 710072 China

# 1. Introduction

The development of artificial intelligence (AI) has brought convenience to people's life. The machine learning algorithms behind AI has also affected various industries and fields, causing a wave of technological reforms. The key factor that promotes the development of AI is the available big data, which is the natural characteristic of computational fluid dynamics (CFD). Therefore, many researchers adopt machine learning techniques to process and analyze massive amounts of CFD simulation and flight test/wind tunnel experiment data. Machine learning provides a modular and agile modeling framework that can be tailored to address many challenges in fluid mechanics, such as reduced-order modeling, experimental data processing, shape optimization, turbulence closure modeling, and control[1]. However, machine learning algorithms cannot be applied perfectly in the field of CFD. Here we are mainly concerned with convolutional neural network (CNN)[2]. As one of the most advancing and widely used methods in machine learning, CNN has achieved great success in the field of image recognition[3], image segmentation[4] and image super-resolution[5]. Nowadays, CNN is adopted widely in image processing competitions. Khan et al.[6] reviewed the origin, innovation, development and achievements of CNN. The hierarchical feature extraction ability of CNN emulates the deep and layered learning process of the neocortex in the human brain, which dynamically learns features from the raw data[7]. Therefore, it can enlighten the understanding and knowledge of complex flow mechanisms by applying CNN to the field of CFD. In addition, higher modeling accuracy is expected to be achieved with the help of CNN.

The traditional convolutional operation is designed to process image pixel data. Therefore, CNN can only be applied directly to the flow-field data on a uniform Cartesian mesh. At present, the application of CNN in the field of CFD is still being explored, and most researchers adopt CNN in the following three strategies:

1. *The local point-to-point mapping is constructed by fully-connected neural network (FNN), which can be regarded as CNN with all filters of size equal to $1\times1$. With* the tensor bases constructed from local flow-field information as input features, Ling et al.[8] employed a deep FNN to predict the Reynolds stress. In addition, in order to achieve efficient aerodynamic shape optimization, Xu et al.[9] constructed the mapping from local flow-field information to local adjoint vectors by a deep FNN. In this modelling strategy, careful design of input features is required in order to implicitly exploit the features of neighbor points.

2. *The flow-field information is obtained by CFD solver on a uniform Cartesian mesh. Therefore, the traditional convolutional operation can be adopted directly.* In order to reconstruct the high-resolution flow field for two-dimensional homogeneous turbulence, Fukami et al.[10] took the low-resolution velocity or vorticity on a uniform Cartesian mesh as input, and reconstructed the high-resolution data via the proposed skip-connection/multi-scale CNN. Beck et al.[11] studied the turbulence modelling via Residual CNN[12] for large eddy simulation on a uniform Cartesian mesh. Zhou et al.[13] constructed a deep CNN to learn the constitutive relation of steady-state convection-diffusion-reaction partial differential equation, and it provides inspiration for more accurate turbulence modelling. However, the requirement of using a uniform Cartesian mesh cannot be met by most industrial applications.

3. *The flow-field information is projected onto a uniform Cartesian mesh for convolutional operation.* In order to predict the unsteady flow field, Han et al.[14] firstly obtained the flow-field data on a uniform Cartesian mesh by projecting from data on a structured mesh. Then a convolutional long short-term memory network[15, 16] is trained on the projected sample data and tested by predicting flow fields in future time steps. However, this strategy has a main problem on the scale of Cartesian mesh, which results in the inaccurate prediction of complex flow nearby the airfoil and thus the incorrect aerodynamic coefficients.

Although CNN can effectively exploit the features of neighbor nodes, the traditional convolutional operation cannot be adopted directly on the most CFD data which calculated on the structured/unstructured mesh. Therefore, the graph convolutional network (GCN), whose convolutional operation is designed for unstructured data, is introduced.

The GCN approaches can be fell into two categories, spectral-based and spatial-based. Bruna et al.[17] firstly proposed projecting graph data into spectral domain through the graph Fourier transformation, and the graph convolutional operation is defined on the spectral domain. Due to the expensive calculation cost, the later works[18, 19] focus on the approximations of spectral-based GCN. However, the learned filters depend on the graph structure, which means the spectral-based GCNs cannot be directly applied to graph data with different structure[20]. In the field of spatial-based GCN, the data structures of image can be a special instance of graphs by regarding each pixel as a node. For a general graph, the spatial-based GCN produces an output of node by aggregating the features of the node and its neighbors. The developments of spatial-based GCN are introduced detailly in the survey paper by Wu el al.[21], and spatial-based GCNs can be divided into two main categories, recurrent-based and composition-based. Different from the main categories, MoNet[22] purposed a new framework to conclude several previous works[18, 23-26]. Under MoNet framework, the weights for features aggregation are determined by the weight function and the relative position between the center node and its neighbors. Similar to the weights sharing in CNN, MoNet framework share the parameters of weight function, which enables the capability of learning graph data with different structure.

Under MoNet framework, the unstructured convolutional neural network (UCNN) is proposed, which aggregates and effectively exploits the neighbour features through the weight function with learnable parameters. UCNN constructed a new form of weight function based on the distribution of flow-field features. Taking the adjoint vector modelling as the research background, the comparison between UCNN and FNN is studied in the case of invariant mesh and variant mesh respectively. In the case of invariant mesh, the modelling capability of UCNN is verified by comparing the prediction for adjoint vector and optimization results with those obtained by FNN in the test incoming state. In the case of variant mesh, the influence of mesh changing on the modelling capability of UCNN is studied through training and validating on data of different mesh.

## 2. Methodology

### 2.1 Forward propagation process

**1) FNN**

Considering the general problem modelling through neural network in CFD, data is usually stored on unstructured/structured mesh. Let $X[k]_i^n$ represents the value of $k^{th}$ neuron on $n^{th}$ layer of $i^{th}$ node. In FNN, the value of $l^{th}$ neuron on $n+1^{th}$ layer of $i^{th}$ node can be calculated as,

$$X[l]_i^{n+1} = \sigma\left(\sum_{k=1}^{K} \omega_{n,l,k}^* X[k]_i^n + b[l]^n\right) \quad (1)$$

where $\sigma$ indicates the activation function, which is *tanh* function in this paper. K is the total number of neurons on $n^{th}$ layer. $b[l]^n$ indicates the bias of $l^{th}$ neuron on $n+1^{th}$ layer. $\omega_{n,l,k}^*$ is the weight connecting the $k^{th}$ neuron on $n^{th}$ layer and the $l^{th}$ neuron on $n+1^{th}$ layer. The connection weights between layers are usually stored in the matrix. As shown in Eq.(1), only the value of current $i^{th}$ node is used to produce the output. Therefore, FNN constructs a point-to-point mapping function on mesh.

**2) UCNN**

Let $X[k]_{i,j}^n$ indicates the value of the $j^{th}$ neighbor node of $i^{th}$ node. Fig. 1 illustrates the forward-propagation schematic of UCNN.

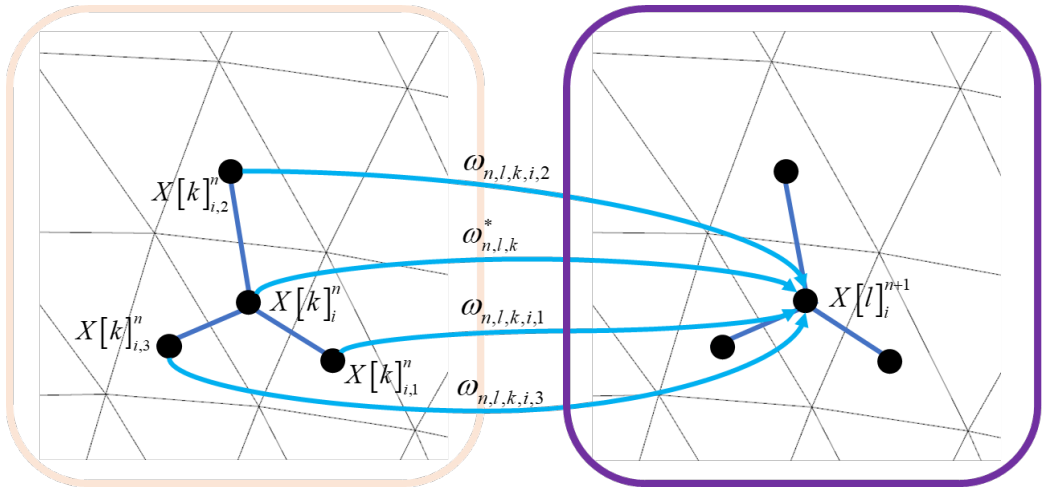

Fig. 1 Forward-propagation schematic of UCNN

In UCNN, $X[l]_i^{n+1}$ is calculated through the neurons on previous layer of current node and its first-order neighbor nodes,

$$X[l]_i^{n+1} = \sigma\left(\sum_{k=1}^{K}\left(\sum_{j=1}^{J}\omega_{n,l,k,i,j}X[k]_{i,j}^n + \omega_{n,l,k}^*X[k]_i^n\right) + b[l]^n\right) \qquad (2)$$

where $\omega_{n,l,k,i,j}$ indicates the weight connecting the $k^{th}$ neuron on $n^{th}$ layer of the $j^{th}$ neighbor node of $i^{th}$ node and the $l^{th}$ neuron on $n+1^{th}$ layer of $i^{th}$ node. Here $\omega$ is the weight function with learnable parameters, which directly influences the modelling capability of UCNN. Once the learnable parameters are determined, the value of $\omega$ is calculated through the relative position between $i^{th}$ node and its $j^{th}$ neighbor node. After careful design and selection of numerical experiments, the weight function is finally determined as,

$$\omega_{n,l,k,i,j} = \beta_{n,l,k} \cdot \sin\left(\gamma_{n,1}\theta_{i,j} + \gamma_{n,2}\right) \cdot \left(\gamma_{n,3}d_{i,j} + \gamma_{n,4}\right) \cdot e^{-d_{i,j}} \qquad (3)$$

where $\theta_{i,j}$ and $d_{i,j}$ indicate the angle and distance of $j^{th}$ neighbor node in the local polar coordinate system, which takes the coordinates of $i^{th}$ node as the origin and calculates the angle counterclockwise from the incoming flow direction. $\gamma$ and $\beta$ are the shared learnable parameters. The neighbor weights of all central nodes are calculated at the same parameters of weight function. However, different weight values will be obtained due to the different relative position in the local polar coordinate system. The proposed weight function can be a product of three parts, the learnable parameter $\beta$, the angle activation function $\sin\left(\gamma_{n,1}\theta_{i,j} + \gamma_{n,2}\right)$ and the distance activation function $\left(\gamma_{n,3}d_{i,j} + \gamma_{n,4}\right) \cdot e^{-d_{i,j}}$. $\beta$ improves the power of weight function. Compared with MoNet which calculate weight values for angle and distance through Gaussian function, the proposed weight function has two main advantages. The first advantage is to put the learnable parameters in the *sin* function and the linear function. Compared with the exponential function, the partial derivatives of *sin* function and the linear function are more stable and robust, which is beneficial to model training and generalization. The second advantage is to adopt a periodic function *sin* with stronger nonlinearity as the activation function for angle. The distance activation function is designed as the product of the linear function part and the exponential part. The exponential part can reduce the influence of $d$ change on the neighbor weights, which makes the model more focused on exploiting the neighbor features in the near-wall area, and appropriately ignores the neighbor features in the far field.

    The idea of UCNN's architecture comes from the Unet[27] as shown in Fig. 2. On the one hand, through skip connection, the low-level features and high-level features can be combined for modelling. On the other hand, it is verified that the skip connection can be beneficial for training such a deep neural network(DNN)[12]. The dotted arrow in Fig. 2 indicates that the low-level features on the left side are copied and concatenated with the high-level features on the right side. The feature number of the layer in UCNN is written on the module, which can be regarded as the neurons number of the layer in FNN. The white arrow represents the unstructured convolutional cell, which is composed of the first-order aggregation, Batch Normalization[28] and activation function *tanh*, as shown in Fig. 3.

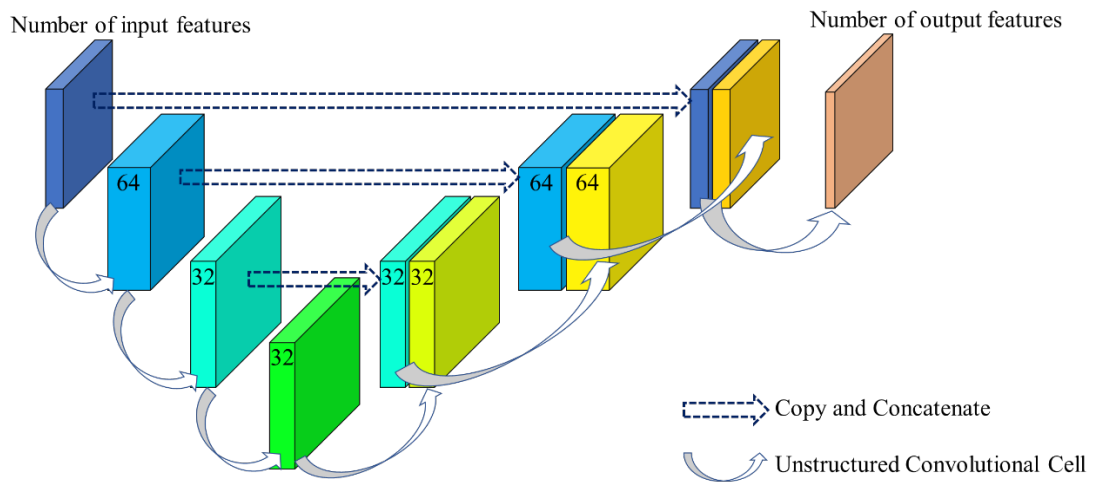

Fig. 2 Architecture of UCNN

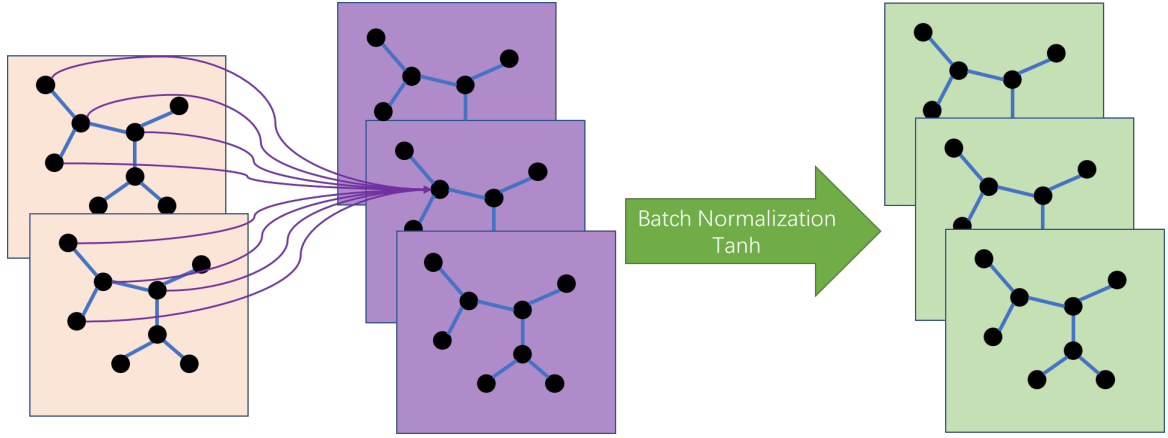

Fig. 3 Framework of unstructured convolutional cell

## 2.2 Efficient parallel implementation on GPU

As mentioned in subsection 2.1, the neighbor weights are calculated based on the relative positions. It is time-consuming and memory-consuming to calculate the neighbor weights of all nodes repeatedly during training. Therefore, the parallel implementation is necessary. It is strongly recommended to run the program on GPU. According to the numerical experiment, the execution speed of program on *RTX2080TI* is at least 50 times faster than *I7-10700*.

Before training, the index matrix $\Gamma$, angle vector $\theta$ and distance vector $d$ are obtained according to the connection relationship and coordinates of the nodes. The pseudo code for generating these vectors and matrix is shown below,

| **Algorithm:** Framework of preprocessor |
|---|
| **Input:** The adjacency matrix $P$ : $P_{i,j}$ indicates the number of $j^{th}$ neighbor node of $i^{th}$ node. The coordinate matrix $u$ : $u_{i,1}$ and $u_{i,2}$ represent $x$ coordinate and $y$ coordinate of $i^{th}$ node respectively. |
| **Output:** Angle vector $\theta$. Distance vector $d$. Index matrix $\Gamma$: The distance and angle between $\Gamma_{i,1}^{th}$ node and $\Gamma_{i,2}^{th}$ node are $d_i$ and $\theta_i$ respectively. |
| **for** $i=1$ **to** total number of nodes **do**     **for** $j=1$ **to** $M_j$ **do** (where $M_j$ is total neighbor number of $i^{th}$ node)         Calculate the distance $t_j$ and angle $a_j$ of $P_{i,j}^{th}$ node in local polar |

> coordinates.
>     **end for**
>     Add $t$, $a$ to $d$ and $\theta$ respectively. Add $[i, P_{i,j}], j=1,2,...,M_j$ to the index matrix $\Gamma$.
> **end for**

After obtaining the required vectors and matrix through the above preprocessor, the calculation process from $X^n$ of $n^{th}$ layer to $X^{n+1}$ of $n+1^{th}$ layer can be simply described as:

1. The activation part of neighbor weights can be calculated through
$$h = \sin(\gamma_{n,1}\theta + \gamma_{n,2}) \bullet (\gamma_{n,3}d + \gamma_{n,4}) \bullet e^{-d} \tag{4}$$

2. The sparse matrix $S$ is defined by taking the $\Gamma$ as the index and $h$ as the value.

3. Finally, $X^{n+1}$ of $n+1^{th}$ layer can be obtained through
$$X^{n+1} = \tanh\left(S \cdot (X^n \cdot \beta) + X^n \cdot \omega^* + b^n\right) \tag{5}$$

For training UCNN on a small batch of data, an efficient approach is inputting all data into UCNN to obtain the full-size output, but the loss is only calculated on the batch. The program is realized in Pytoch[29]. The adopted optimization algorithm and loss function are Adam[30] and mean square error loss (MSELoss) respectively. So far, the efficient parallel implementation for forward propagation and training has been introduced.

Next, the calculation cost of UCNN is studied in *windows10* system with CPU *I7-10700* and GPU *RTX2080TI*. The total number of mesh node for training is 13490. The adopted architecture is shown as Fig. 2. The number of input features and output features are both 4. As shown in Table.1, the total number of learnable parameters and the consumed memory of UCNN are about twice that of FNN.

Table.1. Overview of FNN and UCNN

|      | Total number of learnable parameters | Memory (GB) |
| ---- | ------------------------------------ | ----------- |
| FNN  | 9456                                 | 1.1         |
| UCNN | 18476                                | 2.5         |

By training with different batch size, the training cost and GPU utilization of UCNN and FNN are compared in Table.2 and Table.3. It is obvious that the smaller the batch size, the larger the ratio of GPU utilization of UCNN to that of FNN, which indicates that the parallel efficiency of UCNN is higher. Because the training of UCNN is more complicated, the training cost of UCNN is about 10 times that of FNN even when the batch size is 512.

Table.2. Training cost at different batch size

|      | At batch size=512 | At batch size=8192 |
| ---- | ----------------- | ------------------ |
| FNN  | 10.44 minutes     | 1.42 minutes       |
| UCNN | 103.73 minutes    | 25.51 minutes      |

Table.3. GPU utilization at different batch size

|      | At batch size=512 | At batch size=8192 |
| ---- | ----------------- | ------------------ |
| FNN  | 36%               | 49%                |
| UCNN | 57%               | 62%                |

## 3. Numerical experiments

### 3.1 Problem description

In aerodynamic design, adjoint method[31, 32] is widely used when the number of design variables is large. The calculation cost of adjoint method is irrelevant to the number of design variables, and is merely proportional to the number of objective functions. In many design optimization problems, the number of objective functions is much lower than the number of design variable. Therefore, the adjoint method and gradient-based optimization form a powerful combination[33]. The continuous and discrete adjoint equations can be derived from the continuous and discrete governing equations respectively. The discrete adjoint equation can be solved in higher accuracy. Therefore, the discrete adjoint method[34] is adopted and briefly introduced here.

After the calculation of flow field is completed, we can obtain the objective function $J$ according to the conservative variables $U$ and coordinates of grid points $x_{grid}$,

$$J = J(U, x_{grid}) \tag{6}$$

When the design variables $D$ change, they will affect the value of the objective function $J$ by affecting the conservative variables $U$ and the coordinates of the grid points $x_{grid}$. Therefore, the derivative of the objective function with respect to design variables can be obtained according to chain derivation rule as follows

$$\frac{dJ}{dD_j} = \frac{\partial J}{\partial U}\frac{\partial U}{\partial D_j} + \frac{\partial J}{\partial x_{grid}}\frac{\partial x_{grid}}{\partial D_j}, (j=1...N_D) \tag{7}$$

Equation (7) can be shortened to,

$$\frac{dJ}{dD_j} = \frac{\partial J}{\partial D_j} + \frac{\partial J}{\partial U}\frac{\partial U}{\partial D_j}, (j=1...N_D) \tag{8}$$

where $N_D$ is the number of design variables. $\partial J/\partial D_j$ and $\partial J/\partial U$ on the right-hand side are easy to obtain. However, there is difficulty in calculating $\partial U/\partial D_j$. Therefore, we introduce the adjoint vector $\Lambda$, which can be solved by,

$$\frac{\partial R^T}{\partial U}\Lambda = \frac{\partial J^T}{\partial U} \tag{9}$$

Transposing the above formula and multiplied by $\partial U/\partial D_j$

$$\Lambda^T \frac{\partial R}{\partial U}\frac{\partial U}{\partial D_j} = \frac{\partial J}{\partial U}\frac{\partial U}{\partial D_j} \tag{10}$$

Considering and linearizing the governing equation $R = 0$, then

$$\frac{\partial R}{\partial U}\frac{\partial U}{\partial D_j} = -\frac{\partial R}{\partial D_j} \tag{11}$$

Substituting Eq.(10) and Eq.(11) into Eq.(8), we can get

$$\frac{dJ}{dD_j} = \frac{\partial J}{\partial D_j} - \Lambda^T \frac{\partial R}{\partial D_j}, \ (j=1...N_D) \tag{12}$$

Therefore, once the adjoint vector $\Lambda$ is obtained via Eq.(9), the gradients of all design variables can be obtained via Eq.(12) with negligible cost. The original problem of solving $N_D$ linear equations is transformed into solving only one linear equation. However, the solution of $\Lambda$ is still time consuming. Considering the definition of $\Lambda$ in Eq.(9), $\partial \boldsymbol{R}^T/\partial \boldsymbol{U}$ is a large sparse matrix of size $N_e \times N_e$, and $\partial J^T/\partial \boldsymbol{U}$ is a vector of size $N_e \times 1$. $N_e$ is the total number of grid points, which can be more than $10^4$ in two-dimension case. The calculation cost of $\Lambda$ is approximately equal to once flow computation[35]. In order to improve the adjoint-based optimization efficiency, DNN-based adjoint method is proposed as shown in Fig. 4. DNN is employed to construct a mapping function from the flow-field information $(\rho, u, v, p, ...)$ to $\Lambda$. Then $\hat{\Lambda}$ and the estimated gradients can be obtained via DNN based on the data of $(\rho, u, v, p, ...)$. More information can be referred in Ref[9].

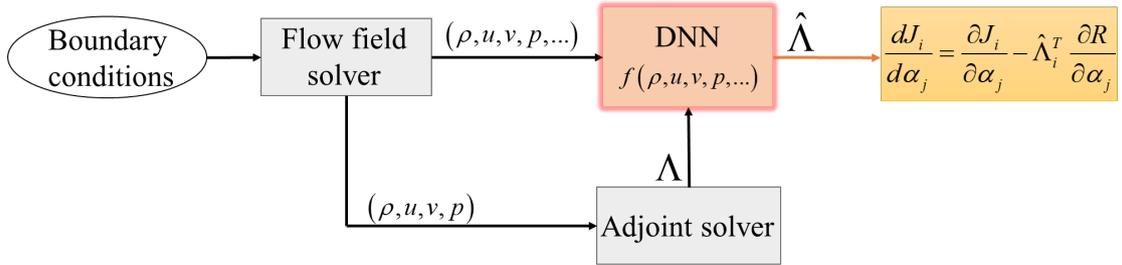

Fig. 4 Flow chart of DNN-Based adjoint method

By training on data at different conditions, DNN can accurately predict $\Lambda$ and achieve the same optimization results as adjoint method. In this paper, we take adjoint vector modelling as the task for UCNN and FNN. The performance of these two approaches will be compared under the same architecture and training settings.

### 3.2 Results

#### 3.2.1 Case of invariant mesh

Firstly, the performance is studied at different $Ma$, different angle of attack ($\alpha$) but invariant mesh. By Latin-hypercube sampling[36] within the range of $Ma = 0.75 \sim 0.85$ and $\alpha = 0° \sim 3°$, 15 incoming states are randomly selected as training samples. 4 validation and 2 test samples are selected to be far away from the training samples, as shown in Fig. 5. The proportions of training, validation and test samples are 71%, 19% and 10% respectively.

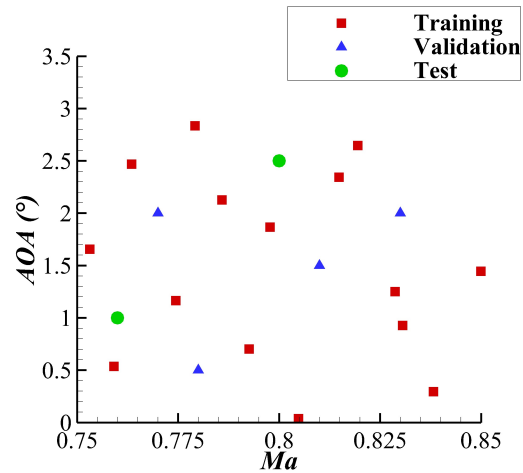

Fig. 5 The distribution of train, validation and test cases

Fig. 6 demonstrates the adopted unstructured mesh with 13490 elements. The data of flow-field features and adjoint vector of NACA0012 airfoil is obtained by solving Euler's equation and discrete adjoint equations at the sampled conditions. The data is stored on the elements. During the training process, the validation loss, which is calculated on the validation set, will not be backpropagated to identify the network weights and biases, but only as reference to prevent overfitting. The test set is not used in the training process.

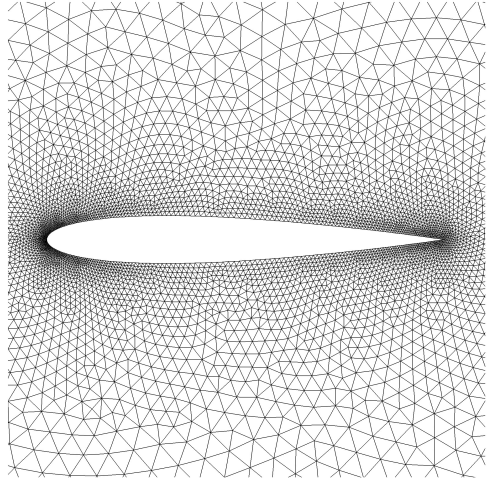

Fig. 6 Unstructured mesh near NACA0012 airfoil

Two combinations of input features are adopted respectively. The first combination is composed of the basic flow features $U=(\rho,u,v,p)$, so we call it *basic features*. The second combination includes basic flow features and its spatial partial derivative $(\partial U/\partial x,\partial U/\partial y)$, so we call it *PD features*. Compared with basic features which represent the local flow-field information, PD features include the local and neighbor flow-field information.

The adopted architecture of FNN is the same as UCNN, which is shown in Fig. 2. Both FNN and UCNN are trained in the following strategy. The training process is divided into two stages. In the first stage, the batch size, the initial learning rate and the number of training epochs is set to $2^{13}$, 0.03 and 800, and the second stage to $2^9$, 0.003 and 500 respectively. During the training process, when no decrease of minimum value of validation loss is seen for 40 training epochs, the learning rate is halved.

Fig. 7 illustrates the convergence of validation loss during training. There appears an oscillation at 800$^{th}$ epoch because the batch size and learning rate are reset. When modelling by basic features, at 133$^{th}$ epoch UCNN achieves the best performance of FNN which is achieved at 1299$^{th}$. When modelling by PD features, at 418$^{th}$ epoch UCNN achieves the best performance of FNN which is achieved at 1296$^{th}$. In addition, the oscillation amplitudes of two approaches are about the same during training. In Fig. 8, the contour images predicted by FNN and UCNN based on basic features are compared with the ground truth. It is obvious that the prediction of UCNN is more accurate.

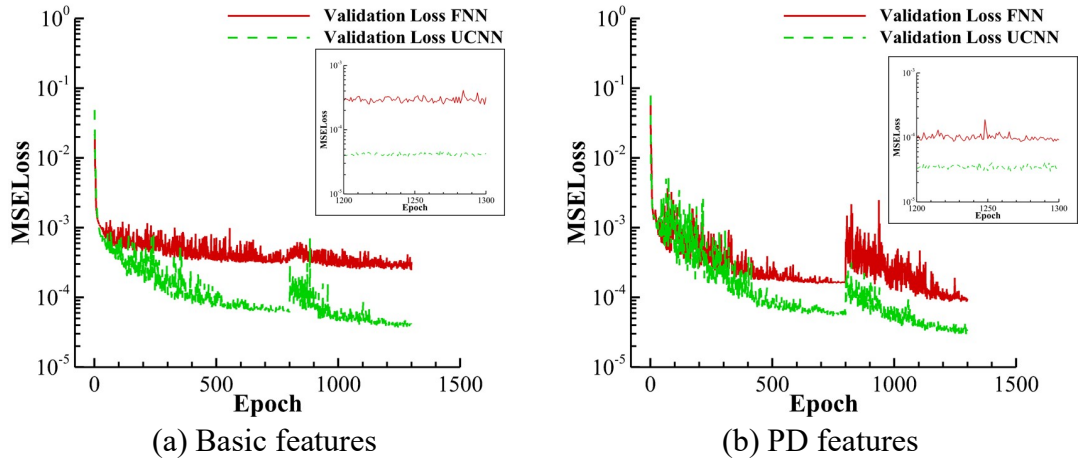

(a) Basic features          (b) PD features

Fig. 7 MSE loss of FNN and UCNN on validation data set during training process

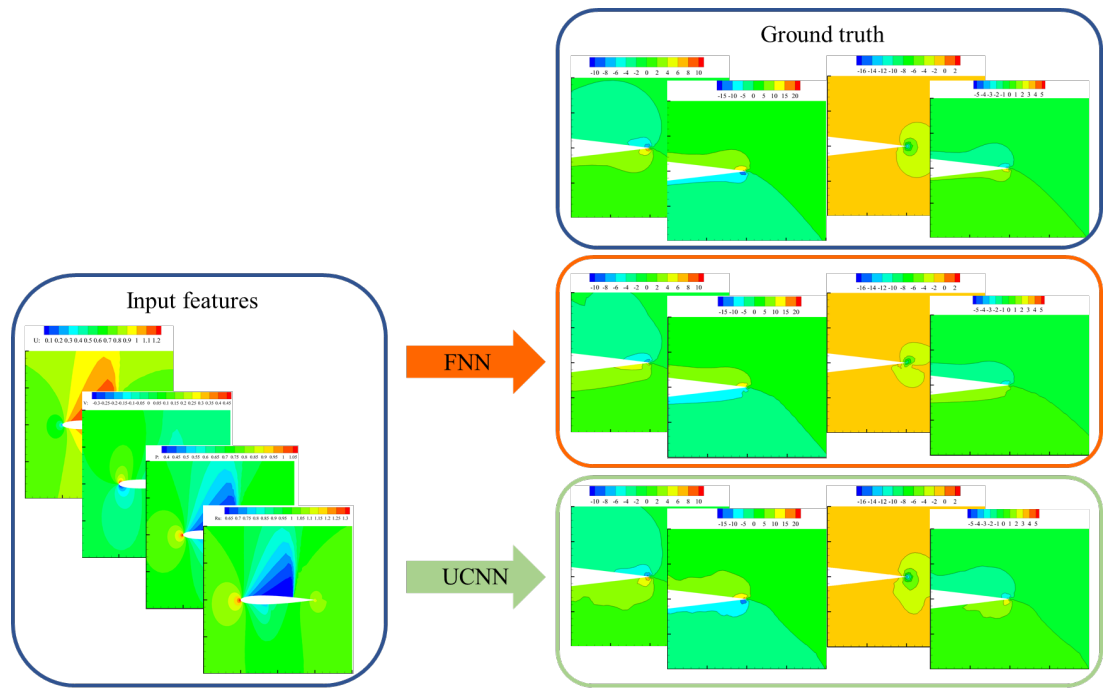

Fig. 8 Prediction of validation sample by FNN and UCNN

In order to reduce the impact of randomness, the training is repeated 5 times. The minimum validation losses and the corresponding train losses are compared in Fig. 9. In general, UCNN achieves a better performance than FNN on the training and validation set. When modelling by basic features (Fig. 9(a)), the minimum validation losses of FNN and UCNN are $2.452 \times 10^{-4}$ and $0.368 \times 10^{-4}$ respectively. The validation loss of FNN is 6.66 times that of UCNN. When modelling by PD features (Fig. 9(b)), the minimum validation losses of FNN and UCNN are $8.646 \times 10^{-5}$ and $2.979 \times 10^{-5}$ respectively. Since the spatial partial derivatives, which denote the neighbor information, are included in the PD features, the validation loss of FNN is only 2.90 times that of UCNN.

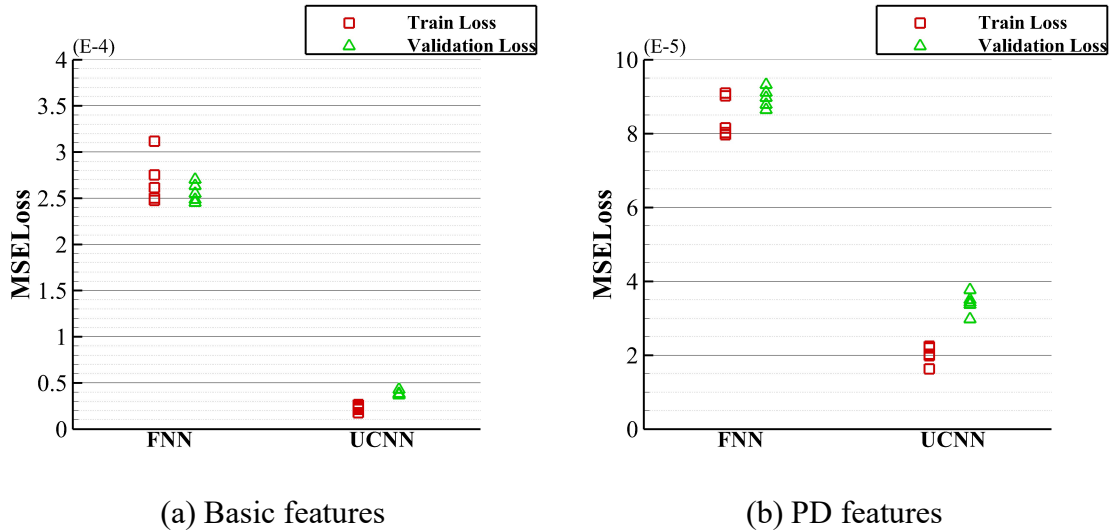

(a) Basic features  (b) PD features

Fig. 9 The MSE loss distribution of FNN and UCNN on train and validation data set

Since modelling by PD features is more accurate, only PD features are adopted in the following aerodynamic shape optimization. The class-shape-transform (CST) methodology[37] is used to parameterize the geometric shape of airfoil. The upper surface and lower surface of the airfoil are both parameterized by 6 CST parameters. The contour images of prediction error on test incoming state 1 ($Ma = 0.80$, $\alpha = 2.5°$) and test incoming state 2 ($Ma = 0.76$, $\alpha = 1.0°$) are shown in Fig. 10(a, b) and Fig. 11(a, b) respectively. The prediction error by FNN is distributed in both near-wall area and far field, while that by UCNN is only distributed in the near-wall area with much smaller value. In Fig. 10(c) and Fig. 11(c), the gradients of drag coefficient with respect to CST parameters predicted by FNN and UCNN are compared with that by traditional adjoint method. The gradients predicted by UCNN is of higher accuracy, especially the gradient for the first CST parameter which has a major influence in the drag reduction.

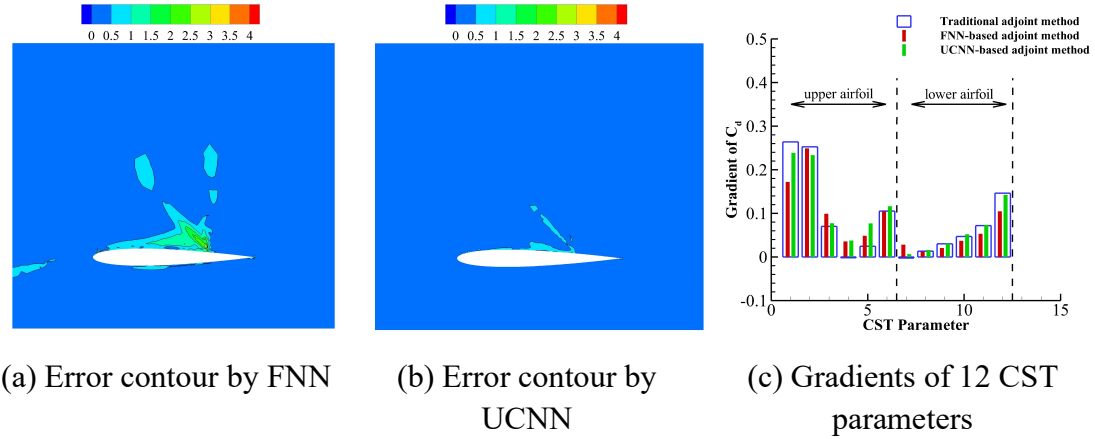

(a) Error contour by FNN  (b) Error contour by UCNN  (c) Gradients of 12 CST parameters

Fig. 10 The error contour and the gradients predicted by FNN and UCNN on test case1

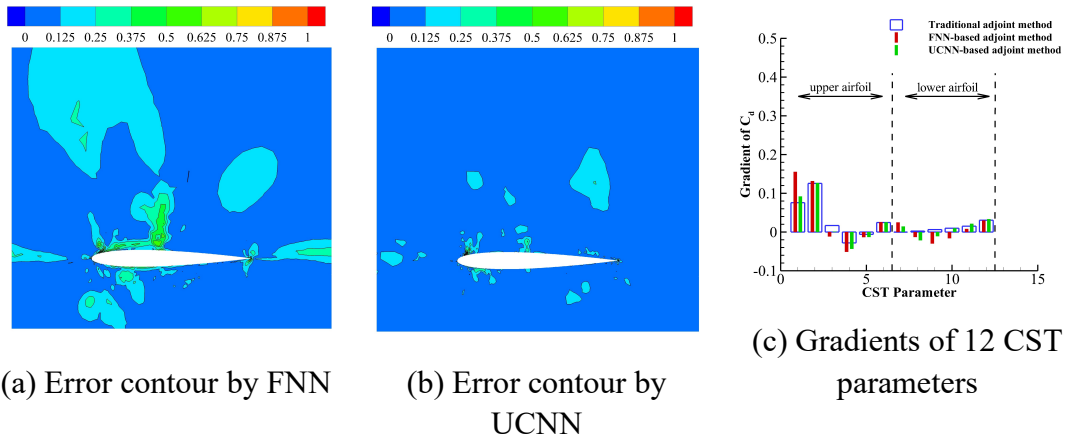

(a) Error contour by FNN  (b) Error contour by UCNN  (c) Gradients of 12 CST parameters

Fig. 11 The error contour and the gradients predicted by FNN and UCNN on test case2

Next, the traditional adjoint method, FNN-based adjoint method and UCNN-based adjoint method are used respectively to guide the drag reduction about NACA0012 airfoil at test incoming state 1($Ma = 0.80$, $\alpha = 2.5°$) in which a certain number of constrains must be satisfied. The lift constrain and area constrain are considered as an equality. The change of CST parameter is limited to be within 30%. The initial lift coefficient $C_{l0}$ is 0.6439, and the initial drag coefficient $C_{d0}$ is 538.9 count. The initial area $Area_0$ is 0.0821. The mathematical model of the optimization

is expressed as follows:

$$\min C_d(\boldsymbol{D})$$
$$\text{s.t.} \quad C_l(\boldsymbol{D}) \geq C_{l0}$$
$$Area(\boldsymbol{D}) \geq Area_0 \tag{13}$$
$$|\Delta CST| \leq 0.3 CST_0$$

where $\boldsymbol{D}$ represents design variable vector, which is CST parameter vector in this paper. The penalty function method is adopted to constrain the optimization problem as follows:

1. $\boldsymbol{D}_0$ is the CST parameter vector of NACA0012 airfoil. $C_{l0}$ of NACA0012 is obtained by the CFD solver. The initial penalty factor $M_1$ and the convergence precision $\varepsilon$ are set to 5 and $1 \times 10^{-10}$, respectively. Let $k=1$.
2. The gradients of $C_d$ with respect to CST parameter are obtained by the traditional adjoint method, FNN-based adjoint method and UCNN-based adjoint method respectively. The gradients of $C_l$ with respect to CST parameter are obtained only by traditional adjoint method. Then the minimum value point $\boldsymbol{D}_k$ can be obtained by solving the following optimization problem,

$$\min f(\boldsymbol{D}_k) = C_d(\boldsymbol{D}_k) + M_k \cdot P(\boldsymbol{D}_k)$$
$$P(\boldsymbol{D}_k) = \Theta_{Cd} + \Theta_{Area} + \Theta_{CST}$$
$$\Theta_{Cd} = \max(C_{l0} - C_l(\boldsymbol{D}_k), 0)^2$$
$$\Theta_{Area} = \max(Area_0 - Area(\boldsymbol{D}_k), 0)^2$$
$$\Theta_{CST} = \max(\boldsymbol{D}_k - ub, 0)^2 + \max(lb - \boldsymbol{D}_k, 0)^2$$

where $ub$ and $lb$ indicate the upper and lower boundary of CST parameter.
3. If $M_k \cdot P(x_k) \leq \varepsilon$, terminate the optimization; else, $M_{k+1} = M_k \cdot 10$ and $k = k+1$, then turn to step 2.

Fig. 12 illustrates the optimal airfoils and NACA0012 airfoil. The optimal airfoils obtained by three methods are almost at the same. The slight difference in the lower surface of optimal airfoils is insensitive to $C_d$. However, the optimal airfoil obtained by UCNN-based adjoint method is closer to that by traditional adjoint method in the leading edge of upper surface, where the shape is most sensitive to $C_d$. Fig. 12 illustrates the $C_d$ iteration of three methods. The traditional adjoint method takes 47 steps to converge while the FNN-based adjoint method and UCNN-based adjoint method take 40 steps and 38 steps respectively. From the $C_d$ iteration, it can be found that FNN-based adjoint method deviate from the traditional adjoint method at 16$^{th}$ step, while this happens at 20$^{th}$ step in UCNN-based adjoint method.

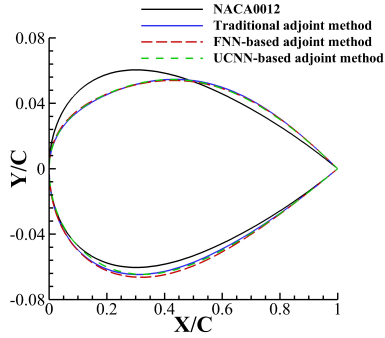 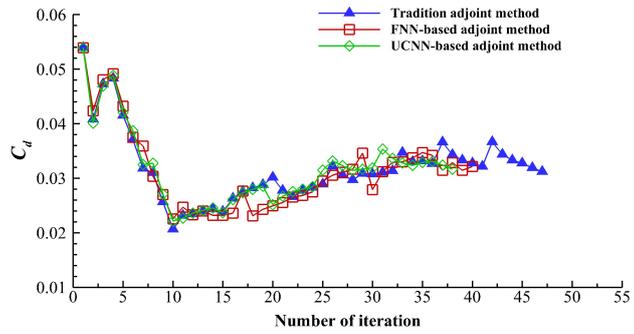

Fig. 12 Comparison of optimal airfoil shape

Fig. 13 $C_d$ iteration during optimization

The aerodynamic coefficients of optimal airfoils are compared in Table.4. The $C_d$ is reduced 40.79% by traditional adjoint method while reduced 39.07% and 39.92% by FNN-based adjoint method and UCNN-based adjoint method respectively, which denotes the superior performance of UCNN.

Table.4. Coefficient of NACA0012 and optimal airfoils.

|  | $C_d$ (1E-04) | $C_l$ | AREA(1E-02) |
| --- | --- | --- | --- |
| NACA0012 | 538.9 | 0.6439 | 8.21 |
| Traditional adjoint method | **319.1** | 0.6465 | 8.26 |
| FNN-based adjoint method | **328.3** | 0.6464 | 8.36 |
| **UCNN-based adjoint method** | **323.8** | 0.6441 | 8.23 |

### 3.2.2 Case of variant mesh

In the previous subsection, the adjoint vector prediction and the drag reduction results obtained by FNN and UCNN are studied. However, the adjoint vector is predicted on invariant mesh and the drag reduction is carried out on deformed mesh, which is insufficient to indicate the influence of mesh changing on the modelling capability of UCNN. Therefore, the influence is further studied in this subsection.

The flow-field features and adjoint vector of NACA0012 are solved on three different unstructured meshes at $Ma = 0.75$ and $\alpha = 2.0°$. The parameters of mesh are shown in Table.5. The main differences when generating mesh 1 and mesh 2 are the number of nodes on the airfoil and the boundary decay. However, unlike the mesh 1 and mesh 2 generated by Delaunay method, mesh 3 is generated by advancing front method, which results in the totally different structure of mesh 3 as shown in Fig. 14. The data obtained by mesh 1 and mesh 2 is selected as the training set while the data obtained by mesh 3 is selected as the validation set.

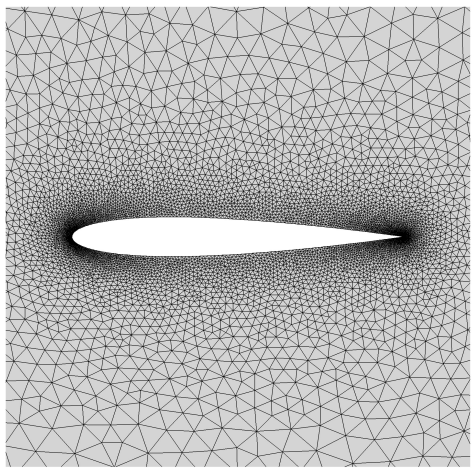
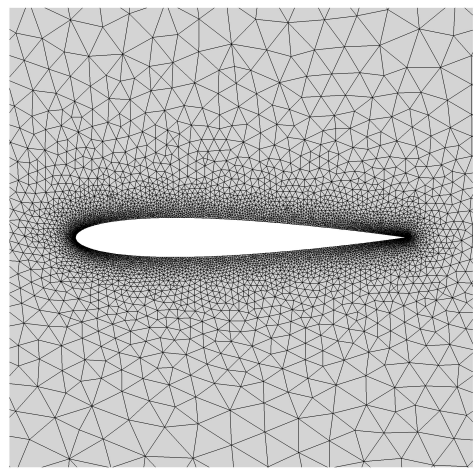

(a) Mesh 1          (b) Mesh 2

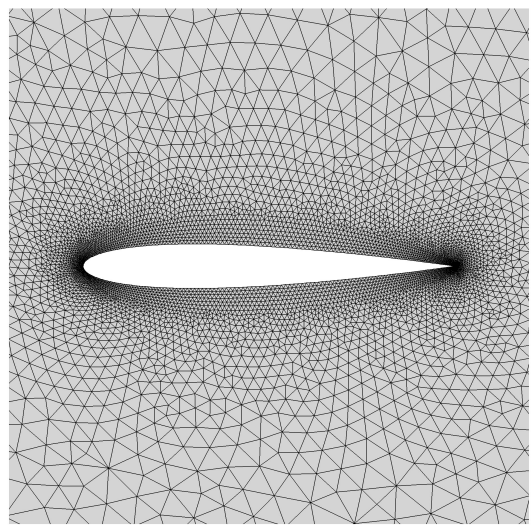

(c) Mesh 3
Fig. 14 Details of meshes

Table.5. Parameters of meshes

|  | Number of nodes on airfoil | Generation type | Boundary decay | Total number of grid nodes | Total number of grid cells |
|---|---|---|---|---|---|
| Mesh1 | 300 | Delaunay | 0.95 | 8522 | 16707 |
| Mesh2 | 400 | Delaunay | 0.9 | 6943 | 13449 |
| Mesh3 | 300 | Advancing Front | 0.95 | 7374 | 14411 |

Since the residual value of the flow field and adjoint vector is converged to $10^{-6}$ under the same boundary conditions, the data of flow-field features and adjoint vector is obtained by three different meshes but indicates the same field. Therefore, FNN can achieve the same performance on training set and validation set, because the local mapping function constructed by FNN is irrelevant to the mesh. By comparing the performance of FNN with that of UCNN, the influence of mesh changing on UCNN can be further studied. There is a numerical error in the solution of spatial partial derivatives due to the different mesh, so only the basic features are used in modelling. The training is repeated 5 times. Fig. 15 illustrates the MSELoss of 5 training on the three meshes. It can be predicted that the MSELoss value of FNN on the three meshes is basically the same. However, UCNN can also achieve the same modeling accuracy on all three meshes after trained on data of mesh 1 and mesh 2. The minimum validation losses by FNN and UCNN are $3.041 \times 10^{-4}$ and $1.218 \times 10^{-4}$ respectively. The validation loss of FNN is 2.50 times that of UCNN. Therefore, UCNN possesses a stronger modelling capability than FNN even in the case of variant mesh.

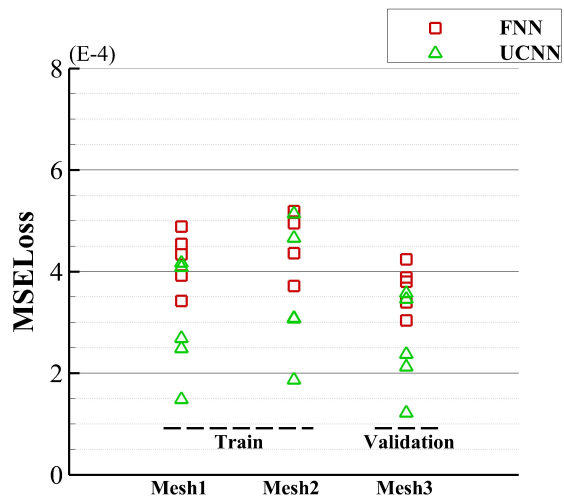

Fig. 15 The distribution of MSE loss on different meshes

## 4. Conclusion

Through aggregating the neighbor features, the modelling capability of UCNN become much stronger than FNN which only uses the local features. Based on the parallel implementation on GPU, the training cost of UCNN is only 10 times that of FNN. Adjoint vector modelling is taken as the task for studying the performance of UCNN. The modelling capabilities of UCNN and FNN are compared in the case of invariant mesh and variant mesh.

In the case of invariant mesh, both UCNN and FNN are trained on data of a dozen of transonic incoming states. The validation loss of FNN is 6.66 times that of UCNN when using the basic features as input, and is 2.90 times when using the PD features as input. An aerodynamic shape optimization is carried out at test incoming state. The optimization results obtained by UCNN-based adjoint method is closer to that by traditional adjoint method. The influence of mesh changing on UCNN is further studied in the case of variant mesh. The data for training and validation is obtained on different meshes. The validation loss of FNN is 2.50 times that of UCNN when using the basic features as input.

The above results indicate that UCNN has a stronger modelling capability and generalization than FNN. Therefore, UCNN is a more powerful choice for modelling task based on structured/unstructured mesh. In the following work, UCNN will be adopted in other fields such as turbulence modelling and flow reconstruction. In addition, the pooling and upsampling operators for UCNN may further improve the modelling capability, and the performance are worth exploring.